\documentclass[10pt]{iopart}

\let\revappendix\appendix
\usepackage{iopams} 
 
\usepackage{amsfonts,bm}
\usepackage{amssymb}
\usepackage{amsmath}
\usepackage{graphicx} 
\usepackage{caption, subcaption}
\usepackage{varwidth}
\usepackage{chngcntr}
\usepackage{comment}
\usepackage{widetext}
\usepackage{cuted}
\usepackage{bm}
\usepackage{tabularx}
\usepackage{color}
\newcolumntype{C}{>{\centering\arraybackslash}X}
\newcolumntype{M}[1]{>{\centering\arraybackslash}m{#1}}
\newcommand{\abs}[1]{\left|#1\right|}
\newcommand{\newc}{\newcommand}
\newc{\beq}{\begin{equation}}
\newc{\eeq}{\end{equation}}
\newc{\beqa}{\begin{eqnarray}}
\newc{\eeqa}{\end{eqnarray}}
\newc{\baln}{\begin{align}}
\newc{\ealn}{\end{align}}
\newc{\balna}{\begin{align*}}
\newc{\ealna}{\end{align*}}
\newc{\mbf}{\mathbf}
\newc{\bsmbl}{\boldsymbol}
\newc{\dg}{\dagger}
\newc{\rgtaw}{\rightarrow}
\newc{\lftaw}{\leftarrow}
\newc{\lrgtaw}{\longrightarrow}
\newc{\llfttaw}{\longrightarrow}
\newc{\mbhe}{{\mbf{\hat{e}}}}
\newc{\dstl}{\displaystyle}
\newc{\jijab}[2]{{J_{i \alpha; j \beta}^{{#1}{#2}}}}
\newc{\jjiba}[2]{{J_{j \beta; i \alpha}^{{#1}{#2}}}}
\newc{\jtab}[2]{{{\tilde{J}}_{\alpha \beta}^{{#1}{#2}}}}
\newc{\jtba}[2]{{{\tilde{J}}_{\beta \alpha}^{{#1}{#2}}}}
\newc{\jtaa}[2]{{{\tilde{J}}_{\alpha \alpha}^{{#1}{#2}}}}
\newc{\sia}[1]{{S}_{i \alpha}^{#1}}
\newc{\sjb}[1]{{S}_{j \beta}^{#1}}
\newc{\mDijab}[2]{{{\mathcal{D}}_{i \alpha; j \beta}^{{#1}{#2}}}}
\newc{\bdg}[1]{b_{{#1}}^\dg}
\newc{\ia}{{i \alpha}}
\newc{\jb}{{j \beta}}
\newc{\sumiajb}{\sum_{\substack{ {i \alpha, ~j \beta} \\ {}}} }
\newc{\sumlims}[2]{\sum\limits_{{#1}}^{#2}}
\newc{\sumliajb}{\sumlims{i\alpha, j\beta}{}}
\newc{\sumlqab}{\sumlims{\mbf{q}, \alpha, \beta}{}}
\newc{\sumsbstck}[2]{{\sum_{\substack{ {{#1}} \\ {{#2}}}} }}
\newc{\mcl}{\mathcal}
\newc{\jtabp}[2]{{{\tilde{J}}_{\alpha \beta^{\prime}}^{{#1}{#2}}}}
\newc{\jtbpa}[2]{{{\tilde{J}}_{\beta^{\prime} \alpha}^{{#1}{#2}}}}

\usepackage{hyperref}
\hypersetup{nolinks=true} 

\begin{document}

\title[Magnon bands in pyrochlore slabs with Heisenberg exchange and anisotropies]{Magnon bands in pyrochlore slabs with Heisenberg exchange and anisotropies}
\author{V. V. Jyothis$^{1,2}$, Bibhabasu Patra$^3$, V. Ravi Chandra$^{1,2}$}
\address{$^1$ School of Physical Sciences, National Institute of Science Education and Research Bhubaneswar, Jatni, Odisha 752050, India}
\address{$^2$ Homi Bhabha National Institute, Training School Complex, Anushaktinagar, Mumbai 400094, India}
\address{$^3$ Dolat Capital Market Pvt. Ltd., 901, Peninsula Park, Veera Desai Industrial Estate, Andheri West, Mumbai 400053, India}
\ead{ravi@niser.ac.in, jyothis.vv@niser.ac.in, patrabibhabasu@gmail.com}

\begin{abstract}

The pyrochlore lattice is a versatile venue to probe the properties
of magnetically ordered states induced or perturbed by anisotropic terms like
the Dzyaloshinskii-Moriya interactions or single-ion anisotropy. Several such
ordered states have been investigated recently as precursors of topological
magnons and the associated surface states. In parallel, there
has been recent progress in growing thin films of magnetic materials with this lattice structure
along high symmetry directions of the lattice. In both cases, an account of the magnetic excitations of relevant
Hamiltonians for finite slabs is a necessary step in the analysis of the physics of
these systems. While the analysis of bulk magnons for these systems is quite common, a direct
evaluation of the magnon spectra in the slab geometry, though required, is less frequently encountered.
We study here magnon bands in the slab geometry for a class of spin models on the pyrochlore lattice with Heisenberg exchange,
Dzyaloshinskii-Moriya interaction and spin-ice anisotropy.
For a range of model parameters, for both ferromagnetic and antiferromagnetic exchange, we compute the classical
ground states for different slab orientations and determine the spin wave excitations above them.
We analyze the ferromagnetic splay phase, the all-in-all-out phase and a coplanar phase
and evaluate magnon dispersions for slabs oriented perpendicular to the $[111]$, $[100]$ and $[110]$ directions.
For all the phases considered, depending on the slab orientation,
magnon band structures can be non-reciprocal and we highlight the differences in the three orientations from
this point-of-view.
Finally, we present details of the surface localized magnons for all the three slab orientations in the phases we study.
For the ferromagnetic splay phase and the all-in-all-out phase we analyze surface states associated
with point degeneracies or nodal lines in the bulk spectrum by computing the magnonic Berry curvature
and Weyl charges or Chern numbers associated with it.

\end{abstract}
%\keywords{Pyrochlore, spin wave spectra, topological magnons}
\submitto{\JPCM}
\ioptwocol
\maketitle

%-------------------------------------------------------------------------------------------------------

\section{Introduction}
\label{intro_section}
A recent development in the field of magnetism has been the entry of
%some materials with 
several magnetically ordered states and the associated magnonic bands into the arena of topological
phases. The analysis of geometric phases \cite{Berry_1984, wilczek_shapere_book} 
of electronic wave functions
on the Brillouin zone, which offers a natural closed parameter space for lattice 
systems, gave rise to revolutionary advances leading to
our present understanding of topological insulators \cite{vanderbilt_book}.
The analogous approach for magnon bands has yielded rich dividends
in the magnetic case as well. Using the study of spin wave band structures
as opposed to electronic ones ordered magnetic materials are now
frequently investigated for topologically non-trivial properties \cite{McClarty_Top_Mag_review, kondo_akagi_katsura_2020_ptep} .
The study of spin waves or magnons, even without the topological gaze, 
is of course indispensable for our understanding of ordered quantum magnets.
They are the elementary excitations above the ground state and measured and evaluated spin wave spectra 
are important tools to extract coupling constants of the underlying lattice model and to understand
their low temperature properties. Unlike electrons, however, magnons 
obey bosonic statistics and this plays a crucial role in the definition and 
analysis of geometric phases for spin wave bands.\\

A notable development which helped initiate this line of study was the prediction \cite{katsura_nagaosa_lee_THE_prl_2010}
and measurement \cite{onose_et_al_science_2010} of the thermal hall effect in a magnetically 
ordered insulator. The non-zero value of the thermal
hall conductivity in $Lu_2 V_2 O_7$ \cite{onose_et_al_science_2010, Ideue_et_al_Lu2V2O7_MHE_prb_2012}, an oxide with the pyrochlore
structure, was shown to be principally due to the magnonic degrees of freedom.
Following the initial prediction \cite{katsura_nagaosa_lee_THE_prl_2010} based on the
presence of non-zero chirality in ordered states of certain lattices with corner-sharing
units, it was soon clear that the transverse thermal conductivity of
magnons is directly related to the (suitably defined) non-vanishing Berry curvature of the 
magnonic band of the underlying Hamiltonian \cite{matsumoto_murakami_MHE_prl_2011, matsumoto_murakami_MHE_prb_2011}.\\
 
There have since been several works studying topological properties of magnons on the pyrochlore lattice.
The initial attempts to theoretically analyze the magnonic thermal hall effect on the pyrochlore
lattice focussed on the $[111]$ orientation of the lattice comprising of alternating
Kagome and triangular lattice planes. The properties of the Dzyaloshinskii-Moriya interaction
on the pyrochlore lattice for an ordered ferromagnetic state along $[111]$
motivated an analysis of the model by confining it to the Kagome planes. 
The resulting studies yielded the magnetic analogues of Chern insulators in two dimensions 
\cite{lifa_zhang_el_al_PRB_2013, mook_henk_mertig_Kagome_2014_PRB, seshadri_sen_kagome_prb_2018, laurell_and_fiete_kagome_PRB_2018}. 
Subsequently, analysis of the magnonic spectra of the full pyrochlore lattice resulted in the discovery 
of analogues of other topological features in three dimensions known for electronic band structures
like Weyl points  \cite{Li_et_al_Weyl_pyrochlore_nature_comm_2016, mook_et_al_prl_2016, su_wang_wang_2017, Jian_Nie_Weyl_2018, magnonic_Weyl_breathing_pyrochlore_2020} 
nodal lines \cite{mook_nodal_lines_2017}, and triple point degeneracies \cite{hwang_et_al_prl_2020}. \\ 

%----------------------------------------------------------------------------------------------------------
%----------------------------------------------------------------------------------------------------------
\begin{figure*}[t]
\centering
\includegraphics[scale=0.5]{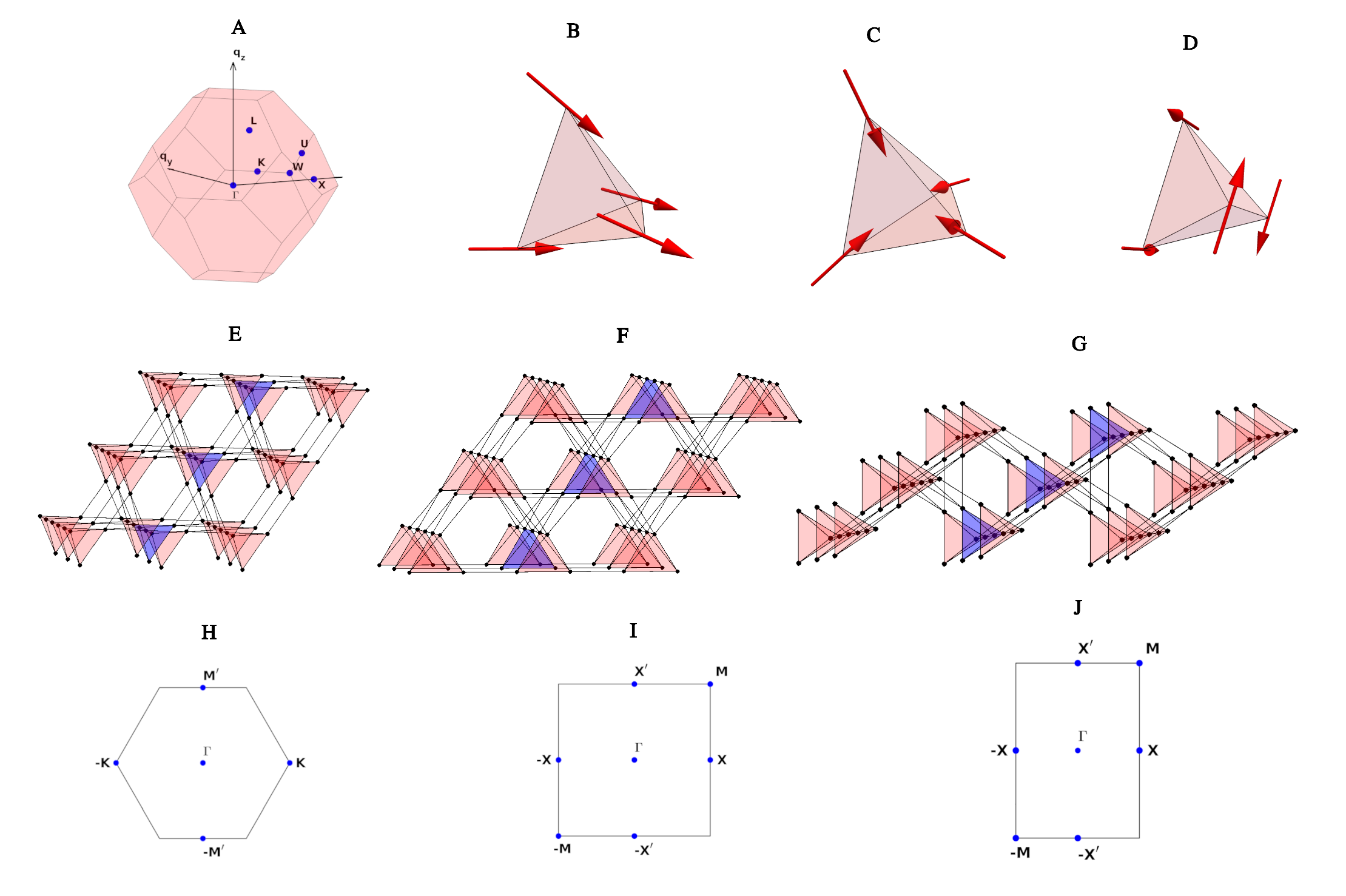}
\caption{{\small{{
A. The Brillouin zone for the pyrochlore lattice. Some high symmetry points mentioned in the
text are depicted in the picture. (B, C, D): The spin orientations in the splay phase, All-in-All-out phase
and the coplanar phase respectively.
(E, F, G): The three pyrochlore slab orientations studied in this work. They are the $[111]$ (E), $[100]$ (F) and $[110]$ (G) 
slabs respectively. For computations in the slab geometry the horizontal directions have periodic boundary conditions. 
A "unit cell" of the effective two dimensional Bravais lattice is depicted using the tetrahedra shaded in blue. 
For clarity a section of the slabs with three layers have been shown, the spectra 
depicted elsewhere in the manuscript are for 60 layers. (H, I, J): The effective two dimensional lattices for the three slab orientations 
are triangular, square and rectangular lattices respectively. The corresponding surface Brillouin zones along with the 
locations of the high symmetry points appearing in subsequent figures are depicted in the bottom row.}}}}
\label{BZ_GS_slabs_SBZ_fig}
\end{figure*}
%----------------------------------------------------------------------------------------------------------
%----------------------------------------------------------------------------------------------------------

Any analysis of topological properties 
of band structures in the bulk involves, almost inevitably, an inspection of the model in the slab 
geometry to probe the existence of edge modes and bulk boundary correspondence. For the pyrochlore lattice,
because of recent advances in preparing and characterizing $[111]$ direction oriented films 
\cite{Thin_film_Weyl_111_Chakhalian_et_al_2021, pyrochlore_thin_films_111_Chakhalian_2021, anil_kumar_et_al_111_films_2022}
such a study in the slab geometry is also relevant on its own. For a set of models on the pyrochlore lattice, studying the 
magnon spectra in the slab geometry is the principal objective of this work. \\

We present a detailed study of the magnon band structures on the pyrochlore lattice
in the slab geometry for a class of models containing Heisenberg exchange, Dzyaloshinskii-Moriya interaction
and spin-ice anisotropy. We begin in Section \ref{prelim_section} by defining the Hamiltonian and present
our evaluation methods and derive some results about the bulk spin waves in the system. 
In Sections \ref{ferro_section} and \ref{afm_section} we present results about the magnon spectra in the
slab geometry for ferromagnetic and antiferromagnetic exchange models respectively. Our conventions for spin wave calculations
and the data depicted in the main text are presented in the Appendices \ref{appendix_SW_theory_details} and \ref{appendix_axes_conventions} respectively. 
%----------------------------------------------------------------------------------------------------------
%----------------------------------------------------------------------------------------------------------
\section{Preliminaries}
\label{prelim_section}
We begin with an introduction of the model Hamiltonian and a brief description
of some aspects of the evaluation of the spin wave spectra. We study the following spin Hamiltonian on the pyrochlore lattice:

\begin{align}
H &= J \sum_{<i\alpha, j\beta>} \mathbf{S}_{i \alpha} \cdot \mathbf{S}_{j \beta} + \sum_{<i\alpha, j\beta>} {\mbf{D}}_{i\alpha, j\beta} \cdot (\mathbf{S}_{i \alpha} \times \mathbf{S}_{j \beta}) \nonumber \\
 &+ K \sum_{{i \alpha}} (\mathbf{S}_{i \alpha}\cdot {{\mbf{\hat{n}}}}_{\alpha})^2 \label{basic_H_test} 
\end{align}

%The three terms are the nearest neighbor Heisenberg exchange, Dzyaloshinskii-Moriya interaction (DMI) and the single-site anisotropy terms respectively.
$\mathbf{S}_{i \alpha}$ are spin-S vector operators at $\mbf{R}_{i \alpha} = \mbf{R}_i + \bm{\alpha}$ = $(\mbf{R}_i, \bm{\alpha})$  where
$\mbf{R}_i$ is a Bravais lattice vector of the FCC lattice with primitive vectors 
${\mbf{a}_i \in \left[(\frac{1}{2}, \frac{1}{2}, 0), (0, \frac{1}{2},\frac{1}{2}), (\frac{1}{2},0,\frac{1}{2}) \right]}$
and $ \bm{\alpha} $ are the four sublattice vectors corresponding to the vertices of the tetrahedron, 
${\bm{\alpha} \in \left[(0,0,0), (\frac{1}{4}, \frac{1}{4}, 0), (0, \frac{1}{4},\frac{1}{4}), (\frac{1}{4},0,\frac{1}{4}) \right]}$.
The Heisenberg exchange $J$ is set to $-1$ or $+1$ for the models studied below for ferromagnetic and
antiferromagnetic exchanges respectively. All the other coupling constants mentioned are in units of $J$.\\

The second term is the Dzyaloshinskii-Moriya interaction (DMI) term which is a leading correction
to isotropic spin exchange when spin-orbit interactions are accounted for in the underlying
model from which the effective spin model is derived \cite{Dzyaloshinsky_1958, Moriya_DMI_prl, Moriya_DMI_Phys_Rev}.
The centre of the nearest neighbour bond is not an inversion centre for the pyrochlore lattice,
which is a necessary condition for the DMI to be non-vanishing. For this lattice, the possible DMI vectors ${\mbf{D}}_{i\alpha, j\beta}$
were worked out in \cite{elhajal_et_al_pyrochlore_dm}.
We follow their convention and assume
that $(\frac{-D}{\sqrt{2}}, \frac{D}{\sqrt{2}}, 0)\cdot (\mathbf{S}_{i 1} \times \mathbf{S}_{i, 2})$ is the DM term
for the first two sublattices. The rest of the vectors are determined by lattice symmetries. Both $D>0$ and $D<0$ are allowed
and have been called "direct" and "indirect" DMI terms respectively in \cite{elhajal_et_al_pyrochlore_dm}, a terminology which we retain.
In ferromagnetic materials with the pyrochlore structure DMI has been discussed recently in $Lu_2 V_2 O_7$ \cite{onose_et_al_science_2010}
and $Ho_2 V_2 O_7$ and $ In_2 Mn_2 O_7 $ \cite{onose_et_al_science_2010, Ideue_et_al_Lu2V2O7_MHE_prb_2012}. The strength of
DMI up to $D/J\approx 0.3$ (and of both signs) have been investigated for this series of materials. For antiferromagnetic exchange as well the
pyrochlore iridate family $A_2 Ir_2 O_7$ has been investigated using DMI strengths of a fraction of the exchange \cite{Krempa_Kim_pyrochlore_Iridates_PRB_2012, hwang_et_al_prl_2020}.
Since the strengths of a fraction of the exchange are commonly studied for pyrochlore materials, in this study we have used  
 $D$ values ranging from  $-0.5$ to $+0.5$. \\

The third term in the Hamiltonian is a single-ion anisotropy term with the coefficient $K$.
The local anisotropy axis unit vector for each spin, $\mbf{\hat{n}_{\alpha}}$, points from the site to the center of the tetrahedron.
The addition of this term in the Hamiltonian plays a part in stabilising ground states with non-collinear order on the lattice.
Recently the spin-3/2 material $Li Ga Cr_4 O_8$ \cite{Li_et_al_Weyl_pyrochlore_nature_comm_2016} and the spin-1 material $Na Ca Ni_2 F_7$ \cite{Li_Chen_spin_1_pyrochlore_2018} have been
analysed using the interplay of single-ion anisotropy and other terms in the Hamiltonian as has also been done in some other contexts for spin models on the pyrochlore lattice.
\cite{DFT_K_pyrochlore_oxides_2011, Jian_Nie_Weyl_2018}. 
We choose our range of $K$ to encompass the diverse possible phases instead of a focusing on a particular material.
 For each value of $D$ the value of $K$ is varied from $-2.0|D|$ to $+2.0|D|$, to establish the robustness
of the results on varying $K$ and to provide analysis of the several phases for this Hamiltonian. Representative results
are presented in the figures for a particular value of $D$ and $K$ but the description holds more generally.
The symbols $J, D, K$ all have dimensions of energy and the symbol $S$, wherever it appears, is a dimensionless constant.
%For all the data sets presented in this work we fix the magnitude of $J$ to be $1$ and study the system for $D$ values
%ranging from $-0.5$ to $+0.5$. For each value of $D$ the value of $K$ is varied from $-2.0|D|$ to $+2.0|D|$. Both ferromagnetic
%and antiferromagnetic models are considered. \\

For fully periodic boundary conditions (PBC), for all the parameter regimes studied, the classical ground states of Eq. \ref{basic_H_test}
can be found using the Luttinger-Tisza (LT) approach \cite{luttinger_and_tisza_1946}. In effect, we evaluate the
eigenvalues of the Fourier transform of the coupling matrix (${\tilde{{\mathcal{J}}}}_{\alpha; \beta}(\lambda, \mu) (\mbf{q})$, 
see Eq. \ref{sw_ham_real_space} in Appendix \ref{appendix_SW_theory_details} )
for each $\mbf{q}$ value. An eigenvector corresponding to the minimum eigenvalue (henceforth called the LT eigenvector) 
results in a ground state spin configuration which satisfies ${\sumlims{i, \alpha}{}} (\mathbf{S}_{i \alpha})^2 = 4 N$, 
which is called the {{\it{{weak constraint}}}}, since it does not guarantee that the obtained spin configuration 
satisfies the {{\it{{strong constraint}}}} $ (\mathbf{S}_{i \alpha})^2 = 1, \forall~  (i, \alpha)$. We have
written the constraint equations for moments of unit magnitude for simplifying notation. The actual magnitude of 
spin will merely change the overall normalisation of the LT eigenvector. If the LT eigenvector does satisfy the 
stronger constraint the minimisation procedure is complete and we have a valid classical ground state of the Hamiltonian. \\ 

For all the phases analyzed here the ground state configuration obtained from the LT eigenvector 
is a $\mbf{q}=0$ state where each sublattice of a tetrahedral unit points in a specified direction. We mainly analyze three ground state
phases: the ferromagnetic splay phase, the all-in-all-out (AIAO) phase and a coplanar phase, depicted in Fig. \ref{BZ_GS_slabs_SBZ_fig}(B, C, D). 
Some of these ground states have been discussed earlier \cite{Li_Chen_spin_1_pyrochlore_2018} in a study
of this model for spin-1 moments using flavor wave theory.\\

Our primary interest in this work is to probe the magnonic surface state structure associated 
with these phases. We consider terminations of the lattice along three of the high symmetry 
directions namely the $[111]$, $[100]$  and the $[110]$ directions. The lattice with slab normals 
along these three terminations is depicted in Figs. \ref{BZ_GS_slabs_SBZ_fig}(E, F, G). For computations in the slab geometry
only the two horizontal directions have periodic boundary conditions and hence form an effective two dimensional
Bravais lattice. 
%The slab normals in Figs. \ref{BZ_GS_slabs_SBZ_fig}(E, F G) are along $[111]$, $[100]$ and $[110]$ respectively. 
The $[111]$, $[100]$ and the $[110]$ terminations of the pyrochlore lattice result in effective 
two dimensional triangular, square and rectangular Bravais lattices with bases respectively. The corresponding surface
Brillouin zones and the locations of the high symmetry points used to present magnon spectra in subsequent
sections are also shown in Fig. \ref{BZ_GS_slabs_SBZ_fig}.
Our conventions for the new axes in terms of which magnon spectra in the slab geometry are 
presented are detailed in \ref{appendix_axes_conventions}.\\

We find that generically, for evaluations in the slab geometry, the Luttinger-Tisza method fails 
to provide a ground state configuration (the strong constraint is violated by the LT eigenvector). 
As a result, in order to evaluate the spin wave spectra in the slab geometry, we evaluate the classical 
ground state using numerical optimization. Assuming translational invariance in the periodic directions,
we express the classical energy of the system as a function ($\mathcal{E} ( {\bsmbl{\Phi}})$)
of ${\bsmbl{\Phi}} = \left ( \{\theta_{l \alpha}\}, \{\phi_{l \alpha}\}  \right )$, 
a set of angular variables containing two angles per spin which fixes its orientation. Here $l$ denotes
the layer index along the slab direction, and $\alpha$ as before specifies the spin within a tetrahedron. Hence for a 
slab with $N_l$ layers the number of "sublattices" in the slab geometry are $4 N_l$ and $ \mathcal{E}({\bsmbl{\Phi}})$
is a function of $8 N_l$ variables.
We numerically minimize $\mathcal{E}({\bsmbl{\Phi}})$ to find the classical ground state configuration in
the slab geometry. We ensure that the minimizing 
spin configurations thus obtained have a 2-norm of the gradient function 
(i.e. $\vert \frac{\partial{\mathcal{E}}}{\partial {\bsmbl{\Phi}}} \vert $ )
which is vanishing up to numerical precision. We also verify that the obtained configuration is a valid
minimum and that the Hessian $H_{ij}=\frac{\partial^2 \mathcal{E}}{\partial_{\Phi_i}\partial_{\Phi_j}}$
does not have any negative eigenvalues. Furthermore, the spin wave Hamiltonian in Fourier space (mentioned below, in Eq. \ref{H_SW_main}) 
derived using our computed classical ground states admits a Cholesky decomposition and hence is positive definite. 
This is in fact a necessary condition for a solution to be a valid starting point for a spin wave evaluation.
We have analyzed results for $N_l=40, 60 \textrm{~and~} 80$ layers. \\

Starting with the computed classical ground states, we study the Hamiltonian (both in the bulk and slab geometry cases) 
in the standard form of linear spin wave theory given below and analyze the magnon spectra:

\beqa \label{H_SW_main}
{H} \approx & {H}_{spin-wave} \hspace{7.0cm} \nonumber \\
= & \dstl{\sumiajb \jijab{3}{3} S^2}   \dstl{-S \sumiajb \jijab{3}{3} \left [\bdg{\ia} b_{\ia} + \bdg{\jb} b_{\jb} \right ]} \hspace{2.0cm} \nonumber \\
 + &  \dstl{{2S \sumliajb \jijab{+}{-} b_{\ia} \bdg{\jb} + \jijab{-}{+} \bdg{\ia} b_{\jb}}} \hspace{2.5cm} \nonumber \\
 + &  \dstl{{2S \sumliajb \jijab{+}{+} b_{\ia} b_{\jb} + \jijab{-}{-} \bdg{\ia} \bdg{\jb}}} \hspace{2.5cm}  \nonumber \\
 =  & ~\mcl{E}_0 + {\sumlims{\mbf{q}}{}}^{\prime} \left [ \begin{array}{cc} ({\mbf{\bdg{\mbf{q}}}})^T & ({\mbf{b_{-\mbf{q}}}})^T \end{array} \right ] \left [ \begin{array}{cc} {\mcl{A}}_{\mbf{q}} & {\mcl{B}}_{\mbf{q}} \\ {\mcl{B}}^{\dg}_{\mbf{q}} & {\mcl{A}}^{T}_{-\mbf{q}} \end{array}\right ] \left [ \begin{array}{c}  \mbf{b_{\mbf{q}}} \\  \mbf{\bdg{-\mbf{q}}}  \end{array} \right ] \hspace{1.0cm}  \\
 =  & ~\mcl{E}_{zero-point} + S~{\sumlims{{\mbf{q}},\alpha}{}}^{} \left[ \omega_{\mbf{q} \alpha}~ f_{\mbf{q}\alpha}^\dg {f}_{\mbf{q}\alpha} \right ] \hspace{2.5cm}  \nonumber
\eeqa

The operators $ \bdg{\ia}, b_{\ia}$ are the Holstein-Primakoff (HP) creation and annihilation operators associated with the spin at the lattice site $\mbf{R}_{i \alpha}$.
$\mbf{b_{\mbf{q}}}$ is a column vector of Fourier transforms of HP annihilation operators and $\mbf{\bdg{\mbf{q}}}$ is a column vector with
the corresponding creation operators in Fourier space. The size of $\mbf{b_{\mbf{q}}}$ or $\mbf{\bdg{\mbf{q}}}$ is the number of sublattices associated
with a given Bravais lattice point. The symbol $ {\sumlims{\mbf{q}}{}}^{\prime}$ indicates that the sum is over each distinct $\mbf{q} \textrm{~and~} \mbf{-q}$ pair.   
A definition of a new set of bosonic operators:
$ \left [ \begin{array}{c}  \mbf{{f}_{\mbf{q}}} \\ \mbf{{f}^{\dg}_{-\mbf{q}}}  \end{array} \right ] = \mbf{\Gamma}_q \left [ \begin{array}{c}  \mbf{{b}_{\mbf{q}}} \\ \mbf{{b}^{\dg}_{-\mbf{q}}}  \end{array} \right ]$
and a determination of $ \mbf{\Gamma}_{\mbf{q}}$ is made using a well established diagonalization procedure  \cite{colpa_diagonalisation_1978}. The optimization
mentioned above to evaluate the correct ground states for the slab geometry is critical to correctly implementing
the numerical Bogoliubov transformation outlined in \cite{colpa_diagonalisation_1978}. $\mbf{H}_{\mbf{q}}$, the 
matrix in Eq. \ref{H_SW_main} and referred to from hereon as the {{\it{dynamical matrix}}}, needs to be positive definite for the procedure and this usually is not the case for inaccurate evaluations of the slab geometry ground state.
The constants appearing in Eq. \ref{H_SW_main} are functions of the original coupling constants in Eq. \ref{basic_H_test} and the orientations of the moments 
in the classical ground state. The relevant Fourier transform conventions, expressions for the various constants, matrices etc appearing in Eq. \ref{H_SW_main} 
are presented in detail in \ref{appendix_SW_theory_details}.\\

%----------------------------------------------------------------------------------------------------------
%----------------------------------------------------------------------------------------------------------

{{\bf{Spatial inversion and spin wave reciprocity}}}: 
It is necessary to state the properties of the spin wave Hamiltonian under spatial inversion to facilitate
the study of topologically non-trivial properties of bulk magnon bands to be discussed below.
The Hamiltonian in real space mentioned in Eq. \ref{H_SW_main} can
be written in an obvious  matrix form (we have omitted the constant term):
\beq
% \dstl{\sumiajb \jijab{3}{3} S^2} + 
H_{real-space}^{SW} =  \left [ \begin{array}{cc} ({\mbf{\bdg{\mbf{R}}}})^T & ({\mbf{b_{\mbf{R}}}})^T \end{array} \right ] \left [ \begin{array}{cc} {{A}}_{\mbf{}} & {{B}}_{\mbf{}} \\ {{B}}^{\dg}_{\mbf{}} & {{A}}^{T}_{\mbf{}} \end{array}\right ] \left [ \begin{array}{c}  \mbf{b_{\mbf{R}}} \\  \mbf{\bdg{\mbf{R}}} \end{array} \right ]  \label{H_SW_real_sp} 
\eeq  
where we have introduced the real space counterparts of $\mbf{b_{\mbf{q}}}, ~\mbf{\bdg{\mbf{q}}}$. 
For all the phases considered, the matrix shown
above is invariant under permutations of site labels corresponding to spatial inversion if the Hamiltonian 
in Eq. \ref{basic_H_test} has periodic boundary conditions for all primitive lattice directions.
On the pyrochlore lattice, with our convention for the sublattice vectors, 
the site obtained by spatial inversion of $(\mbf{R}, \bm{\alpha})$ is $(-(\mbf{R} + 2\bm{\alpha}), \bm{\alpha})$.
This property of the lattice and the invariance of the matrix in Eq. \ref{H_SW_real_sp} together lead to the following
in Fourier space: 

\beq
 {\mcl{A}}_{\mbf{q}} =  \bm{\Phi}_q^{\dg} ~{\mcl{A}}_{-\mbf{q}} ~\bm{\Phi}_q, ~~~{\mcl{B}}_{\mbf{q}} =  \bm{\Phi}_q^{\dg} ~{\mcl{B}}_{-\mbf{q}} ~\bm{\Phi}_q \label{A_q_B_q_trfm}
\eeq

where $ \bm{\Phi}_q$ is a diagonal matrix with $({\Phi}_q)_{i,j} = \delta_{i,j} \exp(-2~i~\bm{\alpha_i}\cdot{\mbf{q}})$. Defining the matrices:
\beq
I_z = \left [ \begin{array}{cc} {{\mbf{1}}} & {{\mbf{0}}} \\  {{\mbf{0}}} & -{{\mbf{1}}} \end{array} \right ], P_q = \left [ \begin{array}{cc} {\bm{\Phi}_q} & {{\mbf{0}}} \\  {{\mbf{0}}} & {\bm{\Phi}_q} \end{array} \right ], I_x =  \left [ \begin{array}{cc} {{\mbf{0}}} & {{\mbf{1}}} \\  {{\mbf{1}}} & {{\mbf{0}}} \end{array} \right ] \label{mat_defs}
\eeq
it can be shown that the following holds:

\beq
P_q~(I_z \mbf{H}_q)~ P_q^{\dg} = -\left[I_x ~ (I_z \mbf{H}_q)~ I_x   \right]^{*} \label{H_q_recip} 
\eeq
where $*$ denotes complex conjugation. Eq. \ref{H_q_recip}
guarantees that if $\vert \bm{\Psi}_\lambda^{\mbf{q}} \rangle $ is an eigenvector of $I_z \mbf{H}_q$ with an eigenvalue $E_{\lambda}$
then $\left [ I_x P_q \vert \bm{\Psi}_\lambda^{\mbf{q}} \rangle \right ]^{*}$ is an eigenvector of $I_z \mbf{H}_q$ with the eigenvalue  $-E_{\lambda}$.
The positive eigenvalues of the spectrum of $I_z \mbf{H}_{\mbf{q}}$ are the magnon frequencies of $\mbf{q}$ and the
negative of the negative eigenvalues are the magnon frequencies of $-\mbf{q}$. Hence Eq. \ref{H_q_recip} guarantees
that the magnon frequencies obey reciprocity i.e. $\omega_{\mbf{q} \alpha} = \omega_{-\mbf{q} \alpha}$. 
We have derived Eq. \ref{H_q_recip} and the reciprocity condition as a direct consequence of the invariance of $H_{real-space}^{SW}$ 
under permutation of site labels corresponding to spatial inversion. We do 
not impose any further restrictions on the underlying classical ground state other than those required by this invariance. The reciprocity
of the spectrum, which for the bulk case follows from Eq. \ref{H_q_recip}, is not obeyed in general in the slab geometry. Hence in the slab geometry there can be 
regions in the surface Brillouin zone (SBZ) where the spin wave spectrum is 
non-reciprocal. We quantify in this work this non-reciprocity using the quantity 
$\mcl{R}_{\mbf{q}} = 2\times {\sum \limits_{\alpha} } \vert \omega_{{\mbf{q}} \alpha} - \omega_{-{\mbf{q}} \alpha} \vert $.
For any q-vector in the surface Brillouin zone $\mcl{R}_{\mbf{q}}$ is a measure of the deviation from reciprocity over all the bands. 
In the following sections we analyze $\mcl{R}_{\mbf{q}}$ for several phases of the model.\\

%----------------------------------------------------------------------------------------------------------
%----------------------------------------------------------------------------------------------------------
\begin{figure*}[t]
\centering
\includegraphics[scale=0.9]{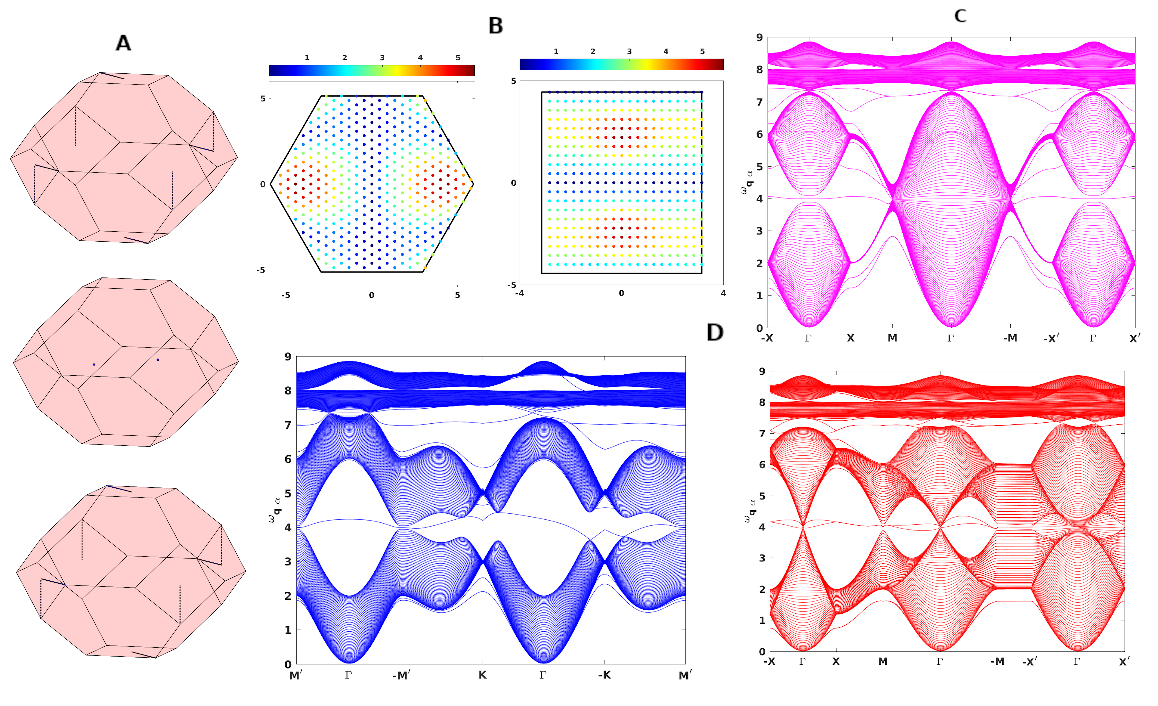}
\caption{{\small{{
Magnons in the ferromagnetic splay phase for $J=-1, D=-K=0.3$: A. Band degeneracy points in the 3D Brillouin zone. 
Top to bottom: minimum gap between second and first band, the third and second band, and the fourth and third band respectively. 
B. $\mcl{R}_{\mbf{q}} = 2\times {\sum_{\alpha}} \vert \omega_{{\mbf{q}} \alpha} - \omega_{-{\mbf{q}} \alpha} \vert $,
$\alpha$ is the band index in the slab geometry. The $[111]$ (left) and $[110]$ (right) geometries have regions of
non-reciprocity. C. Magnon spectrum for a slab of $60$ layers with the $[100]$
termination.  D. Magnon spectrum for a slab of $60$ layers with the $[111]$ termination (Left) and the $[110]$ termination (Right).
}}}}
\label{splay_full_figure}
\end{figure*}
%----------------------------------------------------------------------------------------------------------

%----------------------------------------------------------------------------------------------------------
\begin{figure*}[t]
\centering
\includegraphics[scale=0.9]{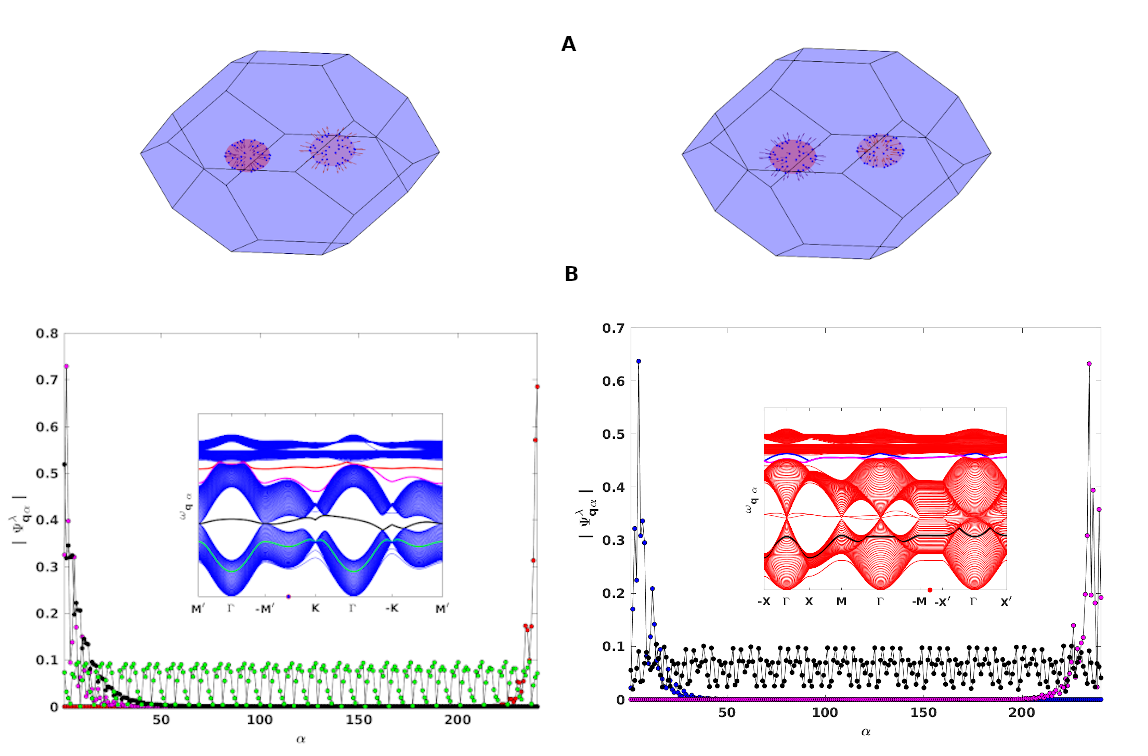}
\caption{{\small{{
A. Berry curvature directions on spheres enclosing the band degeneracies of the second and the third band (see Fig. \ref{splay_full_figure}). The left
picture is for the second band the right one for the third band. B. Depiction of localized and extended modes in the magnon spectrum. The plotted quantity
is the absolute value of the magnon eigenvector of the $\lambda$-th mode at the wave vector $\mbf{q}$ of the SBZ. The insets show in a different
color and thickness the particular magnon mode whose components are being plotted. The colors of the markers in the main plot are chosen to coincide
with the colors of the magnon modes of interest in the inset. Also shown using a dot on the x-axis is the $\mbf{q}$ value for which the analysis
is being done. Data is shown for the $[111]$ (left) and $[110]$ geometry (right). The number of layers in the slab is $60$ and the parameters are $J=-1, D=-K=0.3$. 
}}}}
\label{splay_berry_curvature_and_edge_states}
\end{figure*}

%----------------------------------------------------------------------------------------------------------
%----------------------------------------------------------------------------------------------------------

\begin{figure*}[t]
\centering
\includegraphics[scale=0.9]{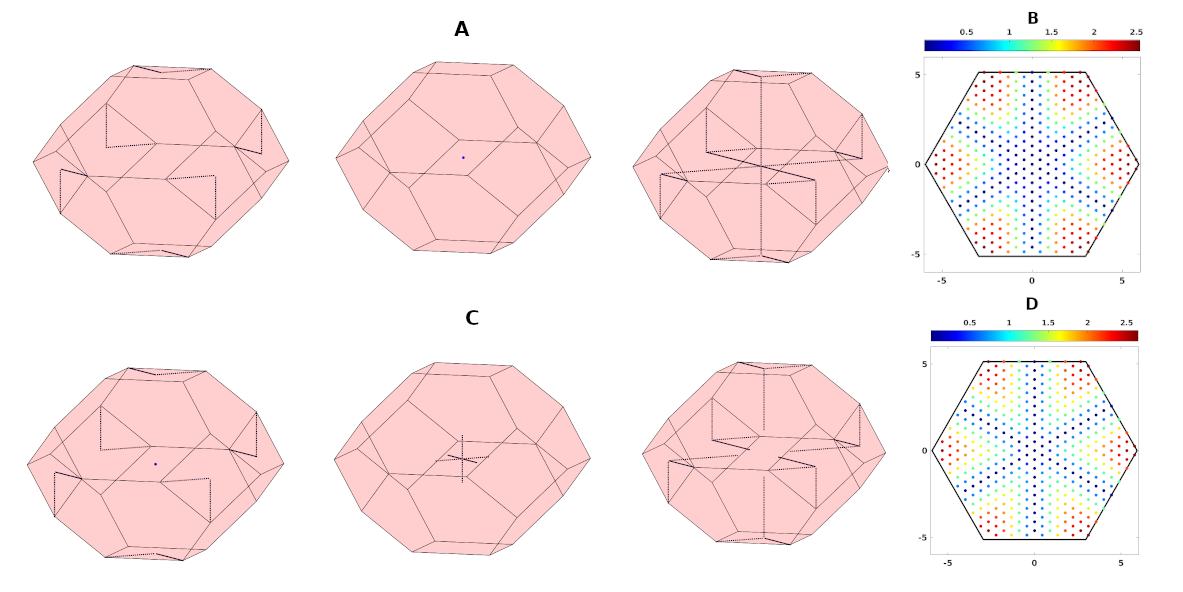}
\caption{{\small{{
(A,C): Degeneracies of bulk bands in the AIAO phase for $J=1.0, D=0.3, K=-0.3$ (A) and $J=1.0, D=0.3, K=0.12$ (C). Points indicated
denote degenerate $\mbf{q}$ vectors for the lowest two bands (left), the second and third bands (middle) and
the top two bands (right). (B, D)  $\mcl{R}_{\mbf{q}} = 2\times {\sum_{\alpha}^{} } \vert \omega_{{\mbf{q}} \alpha} - \omega_{-{\mbf{q}} \alpha} \vert $
for the $[111]$ slab. The $[100]$ and the $[110]$ geometries have reciprocal bands in this phase. The data shown is for $60$ layers.
}}}}
\label{AIAO_gap_and_recip_figure}
\end{figure*}

%----------------------------------------------------------------------------------------------------------
%--------DD EDGE FIGURE
%----------------------------------------------------------------------------------------------------------
 
{{\bf{Magnonic Berry curvature}}}: The analysis of edge magnons involves studying topological aspects
of the bulk spin wave spectrum with periodic boundary conditions and a suitably defined magnonic Berry curvature. 
We employ the definition using which Berry curvature is
commonly computed numerically \cite{Fukui_Hatsugai_Suzuki, vanderbilt_book}.
We calculate the Berry phase associated with a loop enclosing
an elementary plane in the Brillouin zone and the Berry curvature component
perpendicular to the plane is the evaluated Berry phase per unit area. 
Explicitly:

\beqa \label{berry_curvature_def}
\delta \phi = & - \Im \log \left [ \langle \Psi_i \vert I_z \vert  \Psi_j \rangle \langle  \Psi_j \vert I_z \vert  \Psi_k \rangle \langle  \Psi_k \vert I_z  \vert  \Psi_l \rangle \langle  \Psi_l \vert I_z \vert  \Psi_i  \rangle   \right ] \nonumber \\
\Omega_{\perp}^{q} = &  \delta \phi/(\delta A) \hspace{5.0cm}
\label{berry_curvature_definition}
\eeqa

Here, $(i, j, k, l)$ denote the coordinates of an elementary parallelepiped formed using small translations along any two 
of the three reciprocal lattice vector directions. $\Omega_{\perp}^{q}$ is then the component normal to the parallelepiped
at the point $q$, which is at the center. 
The $ \vert  \Psi_{i/j/k/l} \rangle$ are columns of the matrix $\Gamma^{-1}_\mbf{i/j/k/l}$ which are eigenvectors of $I_z \mbf{H}_{i/j/k/l}$, the column index
being determined by the band for which $\Omega_{\perp}^{q}$ is being evaluated.
Hence the underlying matrix whose eigenvectors are used to define the curvature is in general non-Hermitian and requires
a suitable re-definition of the the inner product to the one used in Eq. \ref{berry_curvature_definition}  to define the Berry phase
\cite{shindou_et_al_chiral_2013, matsumoto_shindou_murakami_dipolar_2014, Lein_and_Sato_PRB_2019}.\\

It follows from Eqs. \ref{A_q_B_q_trfm} that $\mbf{H}_{q}$ and  $\mbf{H}_{-q}$ are related as:
\beq
I_z \mbf{H}_{q} = - \left[I_x (I_z \mbf{H}_{-q}) I_x \right]^* 
\label{H_q_H_mq_rel}
\eeq 
An analysis very similar to that after Eq. \ref{H_q_recip} shows that Eq. \ref{H_q_H_mq_rel} implies that $\left [ I_x \vert \bm{\Psi}_\lambda^{\mbf{q}} \rangle \right ]^{*}$ 
is an eigenvector of $I_z \mbf{H}_{-q}$ with the eigenvalue $-E_{\lambda}$, 
if $I_z \mbf{H}_{q} \vert \bm{\Psi}_\lambda^{\mbf{q}} \rangle = E_{\lambda} \vert \bm{\Psi}_\lambda^{\mbf{q}} \rangle$.
Using this relation between the eigenvectors of $I_z \mbf{H}_q$ and the eigenvectors of  $I_z \mbf{H}_{-q}$ with eigenvalues
of the opposite sign, it follows that ${\bm{\Omega}}_{-q, E_\lambda} = -{\bm{\Omega}}_{q, -E_{\lambda}}$. 
In our analysis we have checked that this relation holds to high numerical precision. The interpretation
of certain edge states in the slab geometry to the topological nature of the bulk bands involves
understanding the structure of magnonic Berry curvature close to spectral degeneracies. However, in the case 
of spin wave band structures, because of the bosonic character of the magnons, the Berry curvature over the whole band
contributes to the evaluation of physical quantities like the thermal conductivities (see \cite{zhang_gao_chen_2023_K_xy_review} for a recent review). \\ 
We now present the results of our analysis of spin wave spectra of Eq. \ref{basic_H_test} in the slab geometry for different
phases in three different slab orientations.

%----------------------------------------------------------------------------------------------------------
\section{Ferromagnetic exchange}
\label{ferro_section}

We first consider the case of ferromagnetic exchange interaction $J=-1.0$.
For ferromagnetic (FM) exchange, for the range of DMI strengths studied in this work, 
the term in the Hamiltonian responsible for driving the system away from the purely 
aligned ground state is the spin-ice anisotropy term $K$.
In the absence of $K$, the the DMI term on the pyrochlore lattice 
does not change the energy of the ferromagnetic state with fully aligned spins \cite{onose_et_al_science_2010} and 
hence for all the $D$ values we analyze classical ground state is still a state with all moments aligned.
The continuous degeneracy of this ground state is reduced to a discrete degeneracy of six in the presence of the $K$ term
for both direct and indirect DMI \cite{Li_Chen_spin_1_pyrochlore_2018}. The resulting
classical ground state gives rise to a ferromagnetic splay phase.\\

The splay phase has a ground state spin configuration where all the moments in the tetrahedral unit have a 
common ferromagnetic component and the finite transverse components which add to zero, see Fig. \ref{BZ_GS_slabs_SBZ_fig} (B)
for a schematic representation.  
The Luttinger-Tisza analysis shows that the moments in the tetrahedron to have a form (for the FM component along $\hat{x}$): 
$\alpha_1=\left(a, b, b \right), \alpha_2= \left(a, b, -b \right), \alpha_3= \left(a, -b, -b \right), \alpha_4=\left(a, -b, b \right)$.
The components $a$ and $b$ can be expressed in terms of the coupling constants using a minimization of classical energy
and we omit the detailed expressions.\\

%----------------------------------------------------------------------------------------------------------
In the slab geometry, which is the case of interest for us here, this discrete degeneracy where 
the ferromagnetic components of the moments point along the Cartesian axes remains unchanged 
for the $[111]$ termination of the lattice. 
This happens because the $[111]$ direction
is symmetrically placed relative to all the three Cartesian axes. For our analysis in the slab geometry
we work with the numerically obtained ground states which deep in the bulk have tetrahedra almost in the splay configuration
with the splay axis along the $x$-direction. This choice is convenient because such ground states exist for all the three slab
geometries. As one moves closer to the surfaces the moments can deviate slightly from the splay ground state in 
a manner that depends on the geometry and the number of layers. The salient features of the bulk and the 
corresponding spectrum in the slab geometry for the ferromagnetic splay phase are presented in Fig. \ref{splay_full_figure}.\\
 
On the left we depict those q-values in the 3D Brillouin (BZ) zone at which there are band degeneracies. 
We depict the minimum values of the three successive gaps in the spectrum in the three images of the FCC BZ
from top to bottom. The points shown can either be an exact degeneracy as in the case of the first and the third gap or
a small gap at finite sizes which goes to zero as the lattice dimensions are increased.
The lower two and the upper two bands are seen to be
degenerate along the $X-W$ lines of the Brillouin zone. This is a generic feature for all the phases we analyze
in this work. \\

The second and third band are degenerate at two equal and opposite $q$-vectors as shown. This degeneracy
is a vanishingly small gap at finite lattice sizes which goes to zero in the large lattice size limit.
For the $D=0.3$ and $K=-D$ data presented in Fig. \ref{splay_full_figure} the point degeneracies appear at $(\pm 2.26195, 0, 0)$
for a lattice size of $100 \times 100 \times 100$ Bravais lattice sites and the gap is $\approx 0.018$.
Such point degeneracies in the pyrochlore have been investigated for other Hamiltonians 
earlier \cite{Li_et_al_Weyl_pyrochlore_nature_comm_2016, mook_et_al_prl_2016, su_wang_wang_2017, Jian_Nie_Weyl_2018}
as examples of "magnonic Weyl" points, which are magnetic analogues of the electronic Weyl points \cite{Weyl_Fermi_arc_prb_2011}.   

%----------------------------------------------------------------------------------------------------------
%----------------------------------------------------------------------------------------------------------
\begin{figure*}[t]
\centering
\includegraphics[scale=1.0]{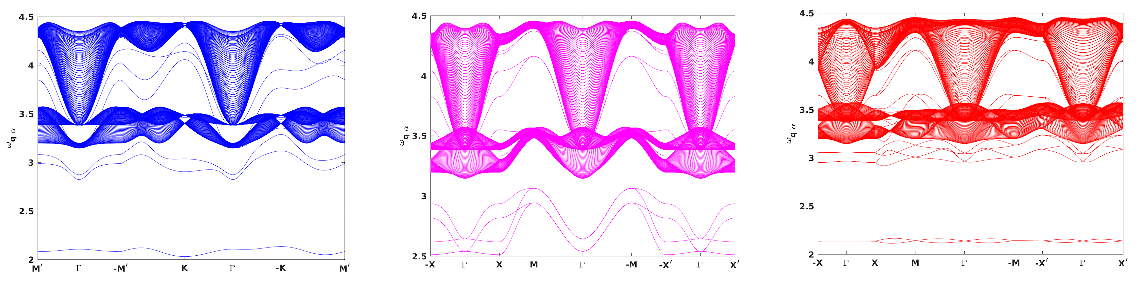}
\caption{{\small{{
The magnon spectrum in the slab geometry in the AIAO phase for $J=1.0, D=0.3, K=-0.3$.
The spectra from left to right are for the $[111]$, $[100]$ and $[110]$ slabs of $60$ layers respectively.
}}}}
\label{AIAO_slab_spectra_figure}
\end{figure*}
%----------------------------------------------------------------------------------------------------------
%----------------------------------------------------------------------------------------------------------

The characteristic feature of such doubly degenerate points is that they can carry quantized topological
charges \cite{Weyl_Fermi_arc_prb_2011}, given by the flux of the Berry curvature over a closed surface 
surrounding the degeneracy ($ C_n = \frac{1}{2\pi}\oint_{S} \bm{\Omega}_{n} \cdot \mbf{dS} $, where $n$ is the band index).  
In Fig. \ref{splay_berry_curvature_and_edge_states} (A),
we depict the the Berry curvature directions on spherical surfaces enclosing the point
degeneracies. We have evaluated $C_n$ numerically for the second and the third bands and found them to be $\approx \pm 1$.
The figures on the left and right in Fig. \ref{splay_berry_curvature_and_edge_states} (A) are for the second and 
third band respectively. We clearly see the reversal of polarities of the curvature for the same point for the 
two bands and for the two $\mbf{q}$-vectors in the same band \cite{Weyl_Fermi_arc_prb_2011}. We have also evaluated Chern numbers on planes
in the Brillouin zone and found that the difference in Chern numbers as one crosses the Weyl points is given
by the quantized charge obtained by evaluation of flux on the sphere surrounding the singularity \cite{Hermanns_Brien_Trebst_Weyl_PRL_2015}.\\

The presence of these quantized charges at point degeneracies and non-zero Chern numbers for planes in the
Brillouin zone are reflected in the surface states arising in the slab geometry depicted in Figs.  \ref{splay_full_figure} (C, D).
There are two kinds of surface states that are clearly visible in the spectrum. The surface states between the first and 
second bands follow from the difference in Chern numbers of the bands involved for any two dimensional section 
of the Brillouin zone. We can also see in Fig. \ref{splay_full_figure} (D) (left) which is for the $[111]$ termination the characteristic surface state
joining the two Weyl points which are along the $M^{\prime}\Gamma(-M^{\prime})$ for the $[111]$ termination.
The spin wave dispersion in the $[111]$ and the $[110]$ terminations have regions
which are non-reciprocal as depicted by the patterns of $\mcl{R}_{\mbf{q}} = 2\times {\sum_{\alpha}^{} } \vert \omega_{{\mbf{q}} \alpha} - \omega_{-{\mbf{q}} \alpha} \vert $
in Fig. \ref{splay_full_figure} (B).  As we will see, such patterns of non-reciprocity exist also for other phases of this model.
The two "magnonic Weyl" points in the bulk BZ are along the $x-axis$ and hence their projections on the surface Brillouin 
zone for the $[100]$ slab are at the origin which implies that the two Weyl points coincide in the surface Brillouin zone. This 
can be seen in Fig. \ref{splay_full_figure} (C), where the only degeneracy between the second and third bands is at the $\Gamma$ point.
Interestingly, $[100]$ slab dispersion is fully reciprocal. This is evident in the surface spectrum for this slab geometry as shown 
in Fig. \ref{splay_full_figure} (C).\\ 

%----------------------------------------------------------------------------------------------------------
In Fig. \ref{splay_berry_curvature_and_edge_states} (B) we depict spatial extent of the several visible mid-gap states
which are clearly marked out in the insets with different colors. We plot, for a slab of $60$ layers and hence $240$ spins the absolute value 
of the components of the eigenstate of $I_z \mbf{H}_{q}$ corresponding to such a state at one $q$-value indicated in the figure (see the marker on the
$x-axis$ of the inset). We also depict for reference one state which is inside the bulk bands for both slab geometries ($[111]$ and $[110]$). 
We note that the states away from the bulk band localize very rapidly 
close to the edges whereas the bulk band eigenvectors (green and black markers for the $[111]$ and $[110]$ slab respectively) have finite contributions
throughout the slab. We note that for both the $[111]$ and the $[110]$ SBZs the locations of the degenerate Weyl $\mbf{q}$-vectors are along
the the y-axis, and hence the regions of non-reciprocal spin waves contain the point degeneracies for the $[110]$ geometry but not for 
the $[111]$ geometry. This can be seen clearly in the dispersion of the mid-gap states in the two geometries. Finally, we note that in the regions where
the spectrum is non-reciprocal, even though $\mcl{R}_{\mbf{q}}$ when it is non-zero is dominated by the mid-gap states the contribution of the 
rest of the states is not negligible. We have illustrated the general properties of this phase using a specific value of $D, K$ in 
Figs. \ref{splay_full_figure}, \ref{splay_berry_curvature_and_edge_states}. These broad features remain the same throughout the phase even 
for other values of $D \in[-0.5, 0.5], K\in[-2 \abs{D}, 2 \abs{D}]$ that we have investigated. 
%----------------------------------------------------------------------------------------------------------
%----------------------------------------------------------------------------------------------------------
%----------------------------------------------------------------------------------------------------------
\begin{figure*}[t]
\centering
\includegraphics[scale=0.6]{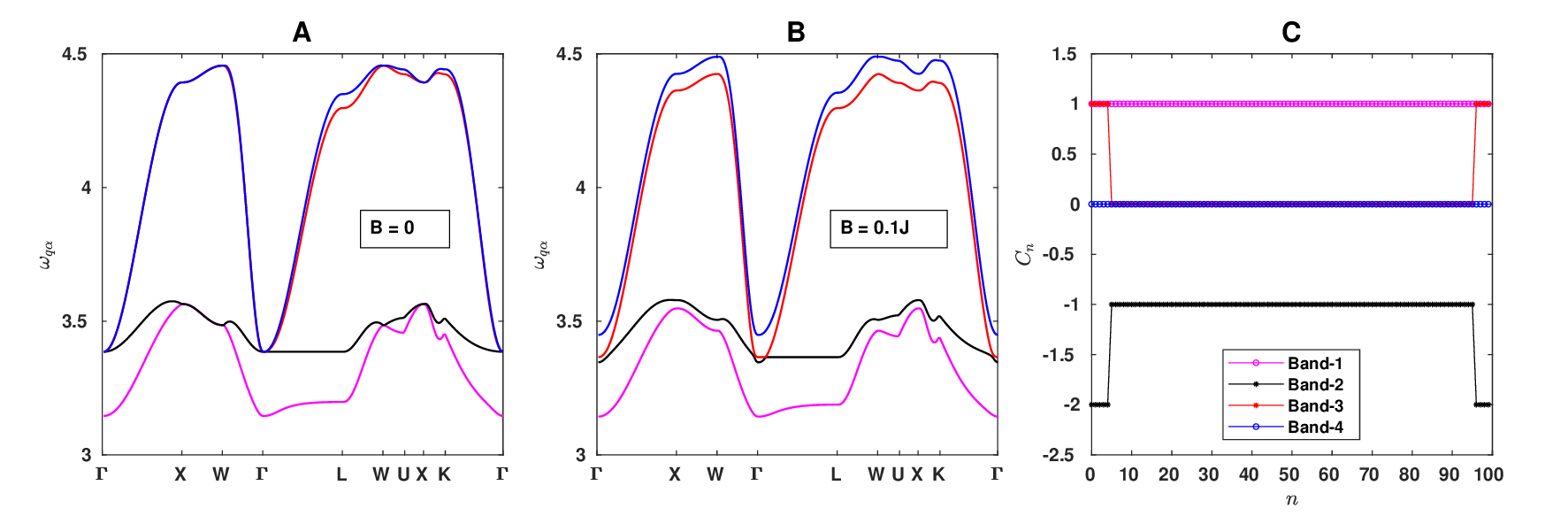}
\caption{{\small{{
A. Bulk bands in the AIAO phase for $J=1.0, D=0.3, K=-0.3$ for zero field and (B) for a small finite field $B=0.1J$ in the $[111]$ direction.
C. Chern numbers of the four bands evaluated on effective two dimensional Brillouin zones obtained by keeping $n=m_3$ in
$\mbf{q} = \sum\limits_{i} \frac{m_i}{N_i} \mbf{K}_i, ~m_i\in [0, N_i-1]$ constant. Here $\mbf{K}_i$ are the basis vectors of
the reciprocal lattice.
The lattice considered has $N_i=100$ for each primitive lattice direction.
}}}}
\label{AIAO_Chern_figure}
\end{figure*}
%----------------------------------------------------------------------------------------------------------
%----------------------------------------------------------------------------------------------------------
%----------------------------------------------------------------------------------------------------------
\section{Antiferromagnetic exchange}
\label{afm_section}
%----------------------------------------------------------------------------------------------------------
%----------------------------------------------------------------------------------------------------------
\subsection{$D>0$: All-in-All-out phase}
The classical ground state and the magnon spectrum in the case of antiferromagnetic exchange 
depends on the sign of DMI. For direct DMI ($D>0$) the classical
ground state is the AIAO state (depicted in Fig. \ref{BZ_GS_slabs_SBZ_fig}(C)) till $\frac{K}{|D|} \approx 1.4$.
In this state all the moments point towards the centre of the tetrahedron of one orientation and away from the centre
of the tetrahedron of the other orientation.
The broad features of the bulk magnon spectrum, which is gapped, show two kinds of triple point degeneracies 
as $K$ is varied. As an illustrative case let us consider $D=0.3$. In the range of $K$ values
studied ($K \in [-2.0\abs{D}, 2.0\abs{D}]$) the system has a triply degenerate $\Gamma$ point with the higher three bands degenerate
from $K=-0.6$ till $K\approx -0.048$. At this $K$ value the $\Gamma$ point is almost four times degenerate and the
lower two and the higher two bands form two sets of doubly degenerate bands respectively. 
Beyond this point the system transitions to a phase with multiple triple point degeneracies. \\

The triple point degeneracy at $\Gamma$ now shifts to the lower three bands and 
three sets of new triple points emerge along the $\Gamma-X$ lines where the upper three bands are degenerate. 
For $K=0$ such a transition between states with different kinds of triple points, as a function of $D$, has been
reported in \cite{hwang_et_al_prl_2020}. Interestingly, in the thin film limit of two or three layers Chern bands 
have been reported in another related model \cite{Laurell_Fiete_2017_prl}.
Beyond $\frac{K}{|D|} \approx 1.4$  the minimum Luttinger-Tisza
eigenvalue is degenerate along continuous lines of the Brillouin zone indicating a 
magnetically disordered quantum ground state. Since we are concerned with ordered states here, 
we do not analyze that regime in this paper.\\ 

%----------------------------------------------------------------------------------------------------------
%----------------------------------------------------------------------------------------------------------

The bulk magnon band degeneracy structure in the two phases is depicted 
in Fig. \ref{AIAO_gap_and_recip_figure} for $D=0.3$, $K=-0.3$ and $K=0.12$. As is clear in the first phase with 
a single triple point at $\Gamma$ the triple point lies at the junction of three nodal
lines along the $\Gamma-X$ directions with the third and fourth bands being degenerate along these nodal 
lines. On the other hand in the second phase with additional triple points along the $\Gamma-X$ directions
these nodal lines split into two parts. On moving away from the triple points on the $\Gamma-X$ line 
the second and the third bands are degenerate along the path to $\Gamma$ and the third and fourth bands are degenerate
along the path to $X$. Thus the nodal lines in the phase with a single triple point at $\Gamma$ are unbroken
throughout the whole zone and connect opposite ends of the Brillouin zone (and hence are closed loops). \\
 
In the second phase with multiple triple points all nodal lines are finite open segments. The lines corresponding to the
the degeneracies of the second and third bands (which include the origin) and those corresponding to the
third and fourth bands meet at the set of triple points along the $\Gamma-X$ directions. This can be seen 
in Fig. \ref{AIAO_gap_and_recip_figure} (C).
%----------------------------------------------------------------------------------------------------------
%----------------------------------------------------------------------------------------------------------
%----------------------------------------------------------------------------------------------------------
\begin{figure*}[t]
\centering
\includegraphics[scale=0.9]{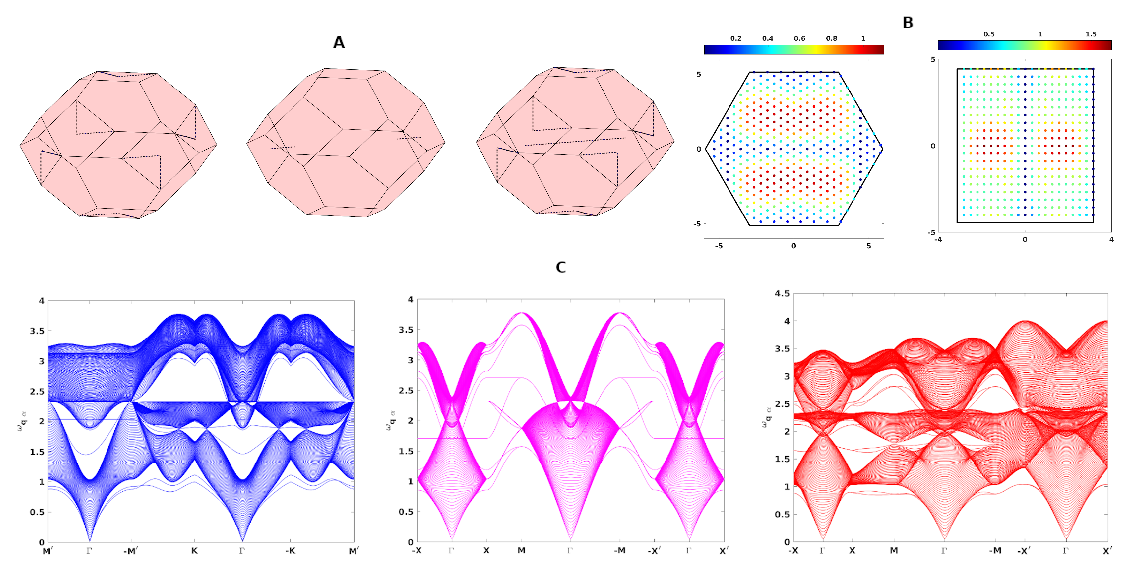}
\caption{{\small{{
A. Degeneracies of bulk bands in the coplanar phase for $J=1.0, D=-0.3, K=0.3$. Points indicated
denote degenerate $\mbf{q}$ vectors for the lowest two bands (left), the second and third bands (middle) and
the top two bands (right). B. $\mcl{R}_{\mbf{q}} = 2\times {\sum_{\alpha}^{} } \vert \omega_{{\mbf{q}} \alpha} - \omega_{-{\mbf{q}} \alpha} \vert $
for the $[111]$ and $[110]$ slab geometries. The $[100]$ geometry spectrum (not shown) is reciprocal over the whole zone.
C. The magnon spectrum in the slab geometry in the coplanar phase for $D=-0.3, K=0.3$. The spectra from left to right are for the $[111]$, $[100]$ and $[110]$
slabs respectively. The data shown is for $60$ layers.
}}}}
\label{DD_full_figure}
\end{figure*}
%----------------------------------------------------------------------------------------------------------
%----------------------------------------------------------------------------------------------------------
%----------------------------------------------------------------------------------------------------------
Also shown in Fig. \ref{AIAO_gap_and_recip_figure} (B, D) are the non-reciprocal regions of the spectrum in the slab geometry.
We find clearly visible regions of non-reciprocal spectrum near the corners of the Brillouin zone for the
$[111]$ termination of the lattice. There is clearly very little difference in the structure of non-reciprocity 
in the phase with a single and multiple triple points. Notably, the regions around the surface Brillouin zone which are
projections of the nodal lines just discussed above have reciprocal magnon spectrum. 
Furthermore, we find that for both the $[100]$ and the $[110]$ terminations the spectrum is
reciprocal. \\

In Fig. \ref{AIAO_slab_spectra_figure} we depict the magnon spectrum in the slab geometry for 
the phase with unbroken nodal lines described above. From left to right are
magnon spectra for $60$ layers for the $[111]$, $[100]$ and $[110]$ terminations respectively.  
For the $[111]$ slab we note the presence of two surface states which cross as the wave vector changes
along the $-M^{\prime} K$ segment of the path. We find that both these states are localized at the two
opposite edges but the location of the surface states is exchanged after the crossing. \\

As in the case of the ferromagnetic splay phase discussed earlier, we now probe the possible
topological origin of the surface states which we find in this phase. Topologically
non-trivial nodal lines are usually characterized by a non-trivial value of the
Berry phase on a small loop encircling the nodal line. On evaluation we find that
the nodal lines we present are not of that category. However, the surface states we present do appear 
for a large range of parameter space and within the gap as denoted in Fig. \ref{AIAO_slab_spectra_figure},
and seem to have their non-trivial topological origin in a closely related Hamiltonian. We find that
the introduction of a small magnetic field lifts the degeneracy associated with the nodal lines and
we only get a point degeneracy at isolated points. For the $[111]$ direction the bulk band structure with
and without the field is shown in Fig. \ref{AIAO_Chern_figure} (A,B). In  Fig. \ref{AIAO_Chern_figure}(C) we
show the variation of the Chern numbers of the four bands evaluated on periodic two dimensional slices
of the three dimensional Brillouin zone. 
Apart from the characteristic changes
in the Chern numbers when the point degeneracy is crossed, we clearly note the different integral
Chern numbers for the second and third bands. The surface states we depict in Fig.  \ref{AIAO_slab_spectra_figure} are
in effect because of these differences in Chern numbers between the second and the third bands on effective 2D Brillouin zones within the three dimensional
Brillouin zone. The reason they appear in a model without the field is essentially because a small field merely acts as a minor 
pertubation, because of the large gap to the excitations. The essential orientation of moments required for the generation of those surface states is already
present in the initial model because of the termination in the slab direction. The small field merely lifts the
degeneracies between bulk bands and produces small gaps which enables us to evaluate the Chern numbers. \\
%----------------------------------------------------------------------------------------------------------
%----------------------------------------------------------------------------------------------------------
The broad features of the surface magnon spectrum is very similar for the phase with finite nodal lines, though the absolute
value of magnon frequencies are different.
We also note the presence of surface states below all the bands (for all the phases). These can
vary widely in dispersion depending on the termination direction and the phase.
%----------------------------------------------------------------------------------------------------------
%----------------------------------------------------------------------------------------------------------
\subsection{$D<0$: Coplanar phase}
For antiferromagnetic exchange and indirect DMI the bulk classical ground state for $K>0$ is a state where all
the spins on a tetrahedron lie on a plane, see for example Fig. \ref{BZ_GS_slabs_SBZ_fig}(D). Such a ground state
is a vector in a two dimensional space of degenerate ground states.
This two dimensional space contains three different coplanar configurations of spins in which the
four spins of the tetrahedron lie in the $x-y$, $y-z$, or the $z-x$ plane. Interestingly, the two dimensional 
space containing these coplanar configurations can be obtained from the ground states discussed 
in a different context \cite{palmer_and_chalker_gs_2000} for the pyrochlore lattice by reversing 
two of the anti-aligned spins. The resulting spin orientation vectors 
(three vectors of $12$ elements) can be seen to span a space of dimension two which contains 
the above mentioned coplanar states.
For our analysis of bulk magnons we work with the coplanar ground state with the spins in the $y-z$ plane, depicted in Fig.  \ref{BZ_GS_slabs_SBZ_fig}(D). \\

In the slab geometry this configuration with deviations in the spin orientations as one moves from from the bulk to the surface
(determined by numerical optimization) yields a classical ground state for all the slab geometries considered. 
Fig. \ref{DD_full_figure} shows the salient features of the magnon spectrum in this phase. The magnon 
spectrum for this phase is gapless. For a given value of $D$ as $K$ is made positive we first observe an almost 
four times degenerate spectrum at the $X$ point which evolves into a triply degenerate point along one of the $\Gamma-X$ lines
in the 3D Brillouin zone. The upper three bands are degenerate at this triply degenerate point. As the value of $K$ increases 
this triply degenerate point travels towards the $\Gamma$ point. Thus we have a triple point at a finite q-vector along 
the $\Gamma X$ path in the Brillouin zone with a partner triple point at the inverted point and both these points are 
connected by a nodal line as in the case with direct DMI. Which of the three $\Gamma X$ lines gets chosen for the nodal line depends on which of the 
coplanar ground states one evaluates the spectrum with. Fig. \ref{DD_full_figure} (A) illustrates what we have described above. \\

Fig. \ref{DD_full_figure} (C) depicts the surface state structure
in this phase for a system of $60$ layers in all the three geometries. We note for the $[111]$ geometry (Fig. \ref{DD_full_figure} (C) (left))
a clear non-dispersive magnon mode around $\omega_{\mbf{q}} \approx 2.03$ which is on the projection of the nodal line on which the third
and fourth band are degenerate. This mode is distinct from the other non-dispersive bulk magnon modes also visible in the plot. 
Such non-dispersive modes are considered to be characteristic features of nodal lines in fermionic bulk band structures under suitable 
conditions \cite{burkov_hook_balents_nodal_semimetals_2011}. In Fig. \ref{DD_full_figure} (B) we depict the pattern of non-reciprocity of 
the magnon spectrum in this phase. We note that unlike in the AIAO phase, the region around the nodal line in the SBZ of the $[111]$ slab 
has a non-reciprocal spectrum. The $[110]$ slab indicates a pattern of non-reciprocity in almost the entire zone. \\ 

We also note the presence of several point degeneracies in Fig. \ref{DD_full_figure} (C) with surface modes traversing between them. 
These point degeneracies do not appear in Fig. \ref{DD_full_figure} (A). That figure depicts the wave vectors with the 
minimum gap for a system size of $100\times100\times100$ unit cells. Hence, an exact degeneracy
if present, is depicted even though the same two bands might also be degenerate in the thermodynamic limit at other isolated points. 
For finte lattice sizes, which are used for numerical analysis, such points can have a small but finite gap.
We note that this occurrence of multiple point degeneracies is also a feature of the $J>0$ model with indirect DMI 
but with $K<0$. For that combination of parameters the system is again in a splay phase.
However, unlike the ferromagnetic splay phase discussed in Section. \ref{ferro_section} the common ferromagnetic
component is much smaller in magnitude than the transverse antiferromagnetic components. This phase also has point degeneracies along
the $\Gamma X$ path. Thus the spin-ice anisotropy on changing from negative to positive values drives a transition
from the antiferromagnetic splay phase to the coplanar phase discussed above. Since we have discussed point degeneracies
and splay phases in Section. \ref{ferro_section}, in the interest of brevity, we have not presented the more involved 
band structures of the antiferromagnetic splay phase in the slab geometry.
Thus, as noted above, the topological structure of the phases with indirect DMI and antiferromagnetic exchange has even richer 
nature than what we have discussed here which needs to be investigated in detail. As in the case of ferromagnetic exchange
the broad details mentioned for specific $D,K$ values remain valid for the full parameter range that we have investigated. 
%----------------------------------------------------------------------------------------------------------
%----------------------------------------------------------------------------------------------------------
\section{Conclusion and outlook}
The interplay of Heisenberg exchange, Dzyaloshinskii-Moriya interaction and the spin-ice anisotropy
on the pyrochlore lattice results in several kinds of ordered states. We have presented here
a detailed analysis of the spin wave spectrum associated with several of those ordered states for
the three commonly studied slab geometries of the lattice. 
Our principal motivation was to study the surface magnon spectra in light of topologically 
non-trivial point degeneracies and nodal lines in the bulk spectra. For the slab sizes 
we have studied we have checked that the surface localized modes change very
little with further increase in the number of layers and hence our results
are a good approximation to the dispersion of the surface states for large lattice sizes
as well. \\ 

The pyrochlore lattice has been a very rich source of topological magnonic phases 
because of its very natural tendency, thanks to anisotropic terms like the
DMI and the spin-ice term, to host non-collinear ordered states which give rise to spin wave 
dynamical matrices with the required properties. While the study of slab geometry is relevant 
for unearthing the surface localized magnons arising from topology of the band structures theoretically, 
for this lattice understanding finite slabs is very relevant experimentally as well. There 
has been recent experimental progress \cite{Thin_film_Weyl_111_Chakhalian_et_al_2021, pyrochlore_thin_films_111_Chakhalian_2021, anil_kumar_et_al_111_films_2022}
in preparing and characterizing $[111]$ oriented thin films with the pyrochlore structure. Furthermore, a proposal for direct detection
of magnonic edge modes by using an electromagnetic drive and a parametric instability has also been reported \cite{ Malz_et_al_Nat_Comm_2019}.
So further rapid progress in the measurement and analysis of magnonic edge states in various systems can be expected in the near future.\\

There are several directions of study starting from our analysis which are immediately
relevant. At the level of linear spin wave theory it would be important to understand the effects of magnetic fields
oriented along different directions relative to a particular slab geometry. In fact, the initial measurements of the thermal 
Hall coefficient for this lattice were indeed carried out with fields oriented along the directions studied in this
work \cite{onose_et_al_science_2010}. Thus the study of the field dependence and the analysis of the thermal Hall physics
in this class of models might legitimately be considered to be the next order of business. Finally, for any
conclusions drawn on the basis of linear spin wave theory, one needs to confront the question of the effects of magnon-magnon
interactions. The analyses along these lines promise to be productive avenues for forthcoming work 
on this interesting class of systems. 
%----------------------------------------------------------------------------------------------------------

\revappendix
\counterwithin{figure}{section}
\section{}
\label{appendix_SW_theory_details}
We provide in this Appendix 
details of our notation used to present spin wave
analysis and related data in the main text. 
This Appendix and \ref{appendix_axes_conventions} are statements of all our conventions
to enable the reader to reproduce the expressions and data sets presented in the text. Though the data
presented in this paper is for the pyrochlore lattice what we say below is
stated in general terms and is true for all systems which satisfy the
specified broad conditions.\\

We consider a generally defined Hamiltonian that is bilinear in spin operators
defined on a Bravais lattice where every Bravais lattice site is occupied by a certain 
number of sublattices:

{{\large{{
\beq
\mbf{H} = \sum_{\substack{ {i \alpha, ~j \beta} \\ {\lambda, \mu}}} {\mathcal{J}}_{i \alpha; j \beta}(\lambda, \mu) ~ S_{i \alpha}^{\lambda} ~  S_{j \beta}^{\mu}  
\label{sw_ham_real_space}
\eeq
}}}}

Here $i, j$ denote the Bravais lattice site indices, $\alpha, \beta$ denote the sublattice indices 
and $\lambda, \mu$ denote Cartesian components of the corresponding spin operators. 
We assume periodic boundary conditions in all the primitive lattice vector directions of the Bravais lattice.
The sums in Eq. \ref{sw_ham_real_space} are all unrestricted. Any double counting resulting from the same
is to be accounted for by choosing the values of the coupling constants ${\mathcal{J}}_{i \alpha; j \beta}(\lambda, \mu)$ to be half of 
the actual values. We note here that $(i \alpha) = (j \beta)$ is allowed, and hence the Hamiltonian can accommodate
single site anisotropy terms. Translation invariance is assumed and hence ${\mathcal{J}}_{i \alpha; j \beta}(\lambda, \mu) = {\mathcal{J}}_{i \alpha + \mbf{R}; j \beta + \mbf{R}}(\lambda, \mu)  $ where $\mbf{R}$ is any Bravais lattice vector.\\ 

Hermiticity of the Hamiltonian demands:
\beq
{\mathcal{J}}_{j \beta; i \alpha}(\mu, \lambda) = \left [ {\mathcal{J}}_{i \alpha; j \beta}(\lambda, \mu) \right ]^{*}
\label{herm_cond_J}
\eeq

The classical ground state about which we develop spin wave theory is assumed to be a state with the translational invariance 
of the underlying Bravais lattice. Hence every sublattice spin has a specific orientation which is the same for all the Bravais lattice sites. 
Many commonly occurring periodic states on Bravais lattices can be represented in this way and also of course non-Bravais
lattices like the honeycomb, Kagome and the pyrochlore lattices.\\ 

We specify the orientation of each sublattice spin $\alpha$ 
using two angles $\displaystyle{ (\theta_\alpha, \phi_\alpha)}$, using standard spherical polar coordinates notation. We define the local
orthogonal right handed axes for each sublattice: 
$\displaystyle{\mbhe_{\alpha}^m}$, $ \displaystyle{\left ({\mbhe_{\alpha}}^1, ~\mbhe_{\alpha}^2,   ~\mbhe_{\alpha}^3 \right ) = \left ({\hat{\theta}}_\alpha, ~{\hat{\phi}}_\alpha, ~{\hat{r}}_\alpha \right ) }  $. 
Thus $\displaystyle{{\mbhe_{\alpha}}^3} $ points along the direction of the sublattice spin in the classical ordered state. The vector
operator of each spin is given by: $ \displaystyle{ {\mbf{S}}_{i \alpha} = \sum_{\lambda} S_{i \alpha}^{\lambda} ~\mbhe_{\lambda} = \sum_{m}  S_{i \alpha}^{m} ~\mbhe_{\alpha}^m} $ where
$\displaystyle{\mbhe_{\lambda} \in (\hat{\mbf{x}},~\hat{\mbf{y}},~\hat{\mbf{z}})} $. Using $\displaystyle {   S_{i \alpha}^{\lambda} =  {\mbf{S}}_{i \alpha} \cdot \mbhe_{\lambda}  } $ in Eq. \ref{sw_ham_real_space}
we arrive at:

{{\large{{
\beq
\mbf{H} = \sum_{\substack{ {i \alpha, j \beta} \\ {m, ~n}}} \mathcal{D}_{i \alpha; j \beta}^{mn} ~ S_{i \alpha}^{m} ~  S_{j \beta}^{n}
\label{sw_ham_real_space_rotated}
\eeq
}}}}

where $m, n \in \left[1, 2, 3  \right] $ and 
\[
%$\displaystyle{\mathcal{D}_{i \alpha; ~j \beta}^{mn} = \sum_{\lambda, \mu}  J_{i \alpha; j \beta}(\lambda, \mu) \left (  {\mbhe_{\alpha}^m} \cdot \mbhe_{\lambda}  \right ) \left (  {\mbhe_{\beta}^n} \cdot \mbhe_{\mu}  \right)}$
\mathcal{D}_{i \alpha; ~j \beta}^{mn} = \sum_{\lambda, \mu}  {\mathcal{J}}_{i \alpha; j \beta}(\lambda, \mu) \left (  {\mbhe_{\alpha}^m} \cdot \mbhe_{\lambda}  \right ) \left (  {\mbhe_{\beta}^n} \cdot \mbhe_{\mu}  \right)
\]

Eq. \ref{herm_cond_J} implies

\beq
\mathcal{D}_{j \beta; ~i \alpha}^{mn} = \left [\mathcal{D}_{i \alpha; ~j \beta}^{nm} \right ]^{*}
\label{herm_cond_D}
\eeq
%%%%%%%%%%%%%%%%%%%%%%%%%%%%%%%%%%%%%%%%%%%%%%%%%%%%%%%%%%%%%%%%%%%%%%%%%%%%%%%%%%%%%%%%%%%%%%%%%%%
%%%%%%%%%%%%%%%%%%%%%%%%%%%%%%%%%%%%%%%%%%%%%%%%%%%%%%%%%%%%%%%%%%%%%%%%%%%%%%%%%%%%%%%%%%%%%%%%%%%
We express the Hamiltonian in terms of the ladder operators $S_{i\alpha}^+ = S_{i\alpha}^1 + i S_{i\alpha}^2 = (S_{i\alpha}^-)^\dg$ :
%\begin{widetext}
\beqa
\mbf{H} = \sum_{\substack{ {i \alpha, ~j \beta} \\ {}}} \left [ \jijab{3}{3} \sia{3} \sjb{3} + \jijab{+}{+} \sia{+} \sjb{+} + \jijab{-}{-} \sia{-} \sjb{-} \right ] & \nonumber \\ 
        + \sum_{\substack{ {i \alpha, ~j \beta} \\ {}}} \left [ \jijab{+}{-} \sia{+} \sjb{-} + \jijab{-}{+} \sia{-} \sjb{+} \right ] & \nonumber \\ 
%\mbf{H} = \sum_{\substack{ {i \alpha, ~j \beta} \\ {}}} \left [ \jijab{3}{3} \sia{3} \sjb{3} + \jijab{+}{+} \sia{+} \sjb{+} + \jijab{-}{-} \sia{-} \sjb{-} + \jijab{+}{-} \sia{+} \sjb{-} + \jijab{-}{+} \sia{-} \sjb{+} \right ] & \nonumber \\ 
% + \sum_{\substack{ {i \alpha, ~j \beta} \\ {}}} \left [  \jijab{+}{3} \sia{+} \sjb{3} + \jijab{3}{+} \sia{3} \sjb{+} + \jijab{-}{3} \sia{-} \sjb{3} + \jijab{3}{-} \sia{3} \sjb{-} \right ]  \hspace{2.35cm} &  
 + \sum_{\substack{ {i \alpha, ~j \beta} \\ {}}} \left [  \jijab{+}{3} \sia{+} \sjb{3} + \jijab{3}{+} \sia{3} \sjb{+} \right ] & \nonumber \\ 
 + \sum_{\substack{ {i \alpha, ~j \beta} \\ {}}} \left [  \jijab{-}{3} \sia{-} \sjb{3} + \jijab{3}{-} \sia{3} \sjb{-} \right ]
\label{sw_ham_ladder_ops}
\eeqa
where the different coefficients are:
\beqa
\jijab{\pm}{\pm} = \frac{1}{4} \left[\mDijab{1}{1} \mp i\mDijab{1}{2}  \mp i\mDijab{2}{1} - \mDijab{2}{2}  \right]~~& \nonumber\\
\jijab{\pm}{\mp} = \frac{1}{4} \left[\mDijab{1}{1} \pm i\mDijab{1}{2}  \mp i\mDijab{2}{1} + \mDijab{2}{2}  \right]~~& \nonumber\\
\jijab{3}{3} = \mDijab{3}{3}~~;~~ \jijab{\pm}{3} = \frac{1}{2} \left[\mDijab{1}{3} \mp i\mDijab{2}{3} \right]~~& \nonumber \\ 
\jijab{3}{\pm} = \frac{1}{2} \left[\mDijab{3}{1} \mp i\mDijab{3}{2} \right] \hspace{1.0cm}
%\jijab{\pm}{\pm} = \frac{1}{4} \left[\mDijab{1}{1} \mp i\mDijab{1}{2}  \mp i\mDijab{2}{1} - \mDijab{2}{2}  \right];~~& \jijab{\pm}{\mp} = \frac{1}{4} \left[\mDijab{1}{1} \pm i\mDijab{1}{2}  \mp i\mDijab{2}{1} + \mDijab{2}{2}  \right];  \nonumber\\
%\jijab{3}{3} = \mDijab{3}{3}~~;~~ \jijab{\pm}{3} = \frac{1}{2} \left[\mDijab{1}{3} \mp i\mDijab{2}{3} \right];~~& \jijab{3}{\pm} = \frac{1}{2} \left[\mDijab{3}{1} \mp i\mDijab{3}{2} \right]; \hspace{3.0cm}
\label{J_consts_real_sp}
\eeqa
An earlier treatment which is similar in spirit and uses Euler angles and rotation matrices can be found in \cite{Haraldsen_spin_rotation_2009}.\\\\
%\end{widetext}

We introduce Holstein-Primakoff bosonic operators $\left(\sia{3} = S - b_{i\alpha}^\dg b_{i\alpha},~ \sia{+} = \sqrt{2S - b_{i\alpha}^\dg b_{i\alpha}}~b_{i\alpha}\right)$ 
in the usual manner and evaluate the effective bosonic Hamiltonian and restrict ourselves to second order terms in the bosonic operators. Minimization of the energy in Eq. \ref{sw_ham_ladder_ops} in the classical limit and demanding that the minimizing spin orientations be along $\displaystyle{{\mbhe_{\alpha}}^3}$ result in all terms that are linear in the bosonic operators being zero.    

We now express the Hamiltonian in Fourier space. Our conventions for Fourier transforms of operators
and various coupling constants appearing in the Hamiltonian are the following:
\beqa
\hat{\mcl{O}}_{q \alpha} & = \frac{1}{\sqrt{N}} \sum\limits_{i} {\hat{\mcl{O}}}_{i\alpha} \exp(-i\mbf{q}\cdot{\mbf{R}}_i) \nonumber \\
\tilde{\mcl{T}}_{\alpha \beta} (\mbf{q}) & = \sum\limits_{{\Delta_{\alpha \beta}}} T({\mbf{\Delta}}_{\alpha \beta}) \exp\left[ -i \mbf{q}\cdot{{\mbf{\Delta}}_{\alpha \beta}} \right]
\label{FT_conv_eqns}
\eeqa
Here, $\dstl{N=\prod_{i} N_i}$ is the number of Bravais lattice sites ($N_i$ sites in each
primitive lattice direction) and $\mbf{q} \in \sum\limits_{i} \frac{m_i}{N_i} \mbf{K}_i,~ m_i\in [0, N_i-1]$, which are all in (or can be brought using lattice translations into) the first Brillouin zone.
$\mcl{O}_{i\alpha}$ is any operator associated with the site $(i\alpha)$. $T({\mbf{\Delta}}_{\alpha \beta})$ can be any coupling 
constant appearing in Eq. \ref{J_consts_real_sp} and $\mbf{\Delta}_{\alpha \beta}$ is the Bravais lattice vector $\mbf{R}_j - \mbf{R}_i$, where we
have used the assumed translation invariance to drop individual indices. 

Using Eqs. \ref{FT_conv_eqns}, the assumption of periodic boundary conditions and standard Fourier identities, the Hamiltonian (Eq. \ref{sw_ham_ladder_ops}) has the following form in Fourier space :
\beq
\label{ham_SW_form_1}
\mbf{H} =   ~\mcl{E}_0 + {\sumlims{\mbf{q}}{}}^{\prime} \left [ \begin{array}{cc} ({\mbf{\bdg{\mbf{q}}}})^T & ({\mbf{b_{-\mbf{q}}}})^T \end{array} \right ] \left [ \begin{array}{cc} {\mcl{A}}_{\mbf{q}} & {\mcl{B}}_{\mbf{q}} \\ {\mcl{B}}^{\dg}_{\mbf{q}} & {\mcl{A}}^{T}_{-\mbf{q}} \end{array}\right ] \left [ \begin{array}{c}  \mbf{b_{\mbf{q}}} \\  \mbf{\bdg{-\mbf{q}}}  \end{array} \right ] \hspace{1.0cm}
\eeq
where, 
%\begin{widetext}
\beqa
%\mcl{E}_0 =  N S(S+1) \sumlims{\alpha \beta}{} \jtab{3}{3}(0) & \nonumber \\
\mcl{E}_0 =  N S(S+1) \sumlims{\alpha \beta}{} \jtab{3}{3}(0) + 2S~ {\sumlims {\mbf{q}, \alpha}{}}^{\prime} \eta_{\mbf{q}} \left [ \jtaa{+}{-}({\mbf{q}}) - \jtaa{-}{+}({\mbf{q}}) \right ] & \nonumber \\
\left [\mcl{A}_{\mbf{q}} \right ]_{\alpha \beta} =  - \eta_{\mbf{q}} S~ {\sumlims{\beta^{\prime}}{}} \left [ \jtabp{3}{3}(0) + \jtbpa{3}{3}(0) \right ] \delta_{\alpha \beta} & \nonumber \\
~~~~~~~~~~~ +  2S~  \eta_{\mbf{q}}  \left [ \jtba{+}{-}({\mbf{q}})  + \jtab{-}{+}({-\mbf{q}}) \right ] \nonumber & \\ 
\left [\mcl{B}_{\mbf{q}} \right ]_{\alpha \beta}  =   ~~2 S~  \eta_{\mbf{q}}  \left [\jtab{-}{-}({-\mbf{q}}) + \jtba{-}{-}({\mbf{q}}) \right  ] 
%\mcl{E}_0 = & N S(S+1) \sumlims{\alpha \beta}{} \jtab{3}{3}(0) + 2S~ {\sumlims {\mbf{q}, \alpha}{}}^{\prime} \eta_{\mbf{q}} \left [ \jtaa{+}{-}({\mbf{q}}) - \jtaa{-}{+}({\mbf{q}}) \right ] \hspace{2.5cm} \nonumber \\
%\left [\mcl{A}_{\mbf{q}} \right ]_{\alpha \beta} = & - \eta_{\mbf{q}} S~ {\sumlims{\beta^{\prime}}{}} \left [ \jtabp{3}{3}(0) + \jtbpa{3}{3}(0) \right ] \delta_{\alpha \beta} + 2S~  \eta_{\mbf{q}}  \left [ \jtba{+}{-}({\mbf{q}})  + \jtab{-}{+}({-\mbf{q}}) \right ] \hspace{1.8cm} \nonumber \\
%\left [\mcl{B}_{\mbf{q}} \right ]_{\alpha \beta}  = & ~~2 S~  \eta_{\mbf{q}}  \left [\jtab{-}{-}({-\mbf{q}}) + \jtba{-}{-}({\mbf{q}}) \right  ] \hspace{6.0cm}
\label{dyn_matrix_elems_eq}
\eeqa
%\end{widetext}

The matrix in Eq. \ref{ham_SW_form_1} is of order $2\times n_{sub}$ where $n_{sub}$ is the number of sublattices. $\mbf{b_{\mbf{q}}}$ and $\mbf{\bdg{\mbf{q}}}$ are column
vectors with entries $\{b_{{\mbf{q}}\alpha}\}$ and $\{\bdg{{\mbf{q}}\alpha}\}$ respectively. The symbol ${\sumlims{q}{}}^{\prime}$ indicates that 
the sum is taken over all distinct pairs $(\mbf{q}, -\mbf{q})$. 
$\eta_{\mbf{q}} $ is ${\frac{1}{2}}$ for those vectors in the Brillouin zone for which $\mbf{q}$ and ${-\mbf{q}}$ differ by a reciprocal
lattice vector and $1$ otherwise. It is straightforward to show using {Eq \ref{herm_cond_D}} and
its implications for lattice Fourier transforms that $\mcl{A}_{\mbf{q}}$ is Hermitian and hence 
the matrix in Eq. \ref{ham_SW_form_1} is Hermitian. We rewrite the Hamiltonian in terms of new bosonic operators 
$ \left [ \begin{array}{c}  \mbf{{f}_{\mbf{q}}} \\ \mbf{{f}^{\dg}_{-\mbf{q}}}  \end{array} \right ] = \mbf{\Gamma}_{\mbf{q}} \left [ \begin{array}{c}  \mbf{{b}_{\mbf{q}}} \\ \mbf{{b}^{\dg}_{-\mbf{q}}}  \end{array} \right ]$:

\beqa
 \mbf{H}~~~~ =  & ~\mcl{E}_{zero-point} + S~ {\sumlims{q,\alpha}{}}^{} \left[ \omega_{\mbf{q} \alpha} f_{\mbf{q}\alpha}^\dg {f}_{\mbf{q}\alpha} \right ] \hspace{2.5cm}  \nonumber\\
\mcl{E}_{zero-point} = & ~~\mcl{E}_0 + S~ {\sumlims{{\mbf{q}},\alpha}{}}^{\prime} \eta_{\mbf{q}}~  \omega_{-\mbf{q}\alpha} \hspace{5.0cm} \nonumber  
\eeqa

Here $\mbf{\Gamma}_{{\mbf{q} }}$ is a transformation matrix from the old to the new bosonic operators, $\mcl{E}_{zero-point}$ is the zero-point energy, and $\omega_{\mbf{q}\alpha}$ 
are the magnon frequencies.
The problem of diagonalisation of a Hamiltonian of the form in Eq. \ref{ham_SW_form_1} by constructing 
the required transformation matrix $\mbf{\Gamma}_{\mbf{q}}$ was considered in detail in \cite{colpa_diagonalisation_1978} and we follow the method outlined there.  

%-----------------------------------------------------------------------------------------------------------
%-----------------------------------------------------------------------------------------------------------
\section{Lattice vectors and axes conventions for slab geometries}
\label{appendix_axes_conventions}

We present here the conventions used to calculate the spin wave spectra in the slab geometry of the
lattice depicted in the main text. Evaluations of the magnon spectra in the slab geometry require the specification of two primitive lattice
vectors to define the effective two dimensional Brillouin zone and a vector to specify how the
2D layers are stacked to build the full three dimensional lattice. The primitive vectors
of the three dimensional lattice are the FCC basis vectors: 
${\mbf{a}_i \in \left[(\frac{1}{2}, \frac{1}{2}, 0), (0, \frac{1}{2},\frac{1}{2}), (\frac{1}{2},0,\frac{1}{2}) \right]}$, where
the side length of the cubic unit cell is set to $1$. For the
three different slab geometries used in this work the effective 2D primitive vectors $\mbf{a}_1^{eff}, \mbf{a}_2^{eff}$ and the stacking vectors $\mbf{a}_S$
are the following:
%\begin{widetext}
\beqa
\textrm{[111]}: \mbf{a}_1^{eff} = \mbf{a}_1 - \mbf{a}_3, ~~\mbf{a}_2^{eff} =  \mbf{a}_2 - \mbf{a}_3 &  \nonumber \\
~~~~~~~~~\mbf{a}_{S} =  \mbf{a}_1  & \nonumber \\
\textrm{[100]}: \mbf{a}_1^{eff} = \mbf{a}_2, ~~\mbf{a}_2^{eff} =  \mbf{a}_3 - \mbf{a}_1  & \nonumber \\ 
~~~~~~~~~\mbf{a}_{S} =  \mbf{a}_1  & \nonumber \\
\textrm{[110]}: \mbf{a}_1^{eff} = -\mbf{a}_1 + \mbf{a}_2 + \mbf{a}_3, ~~\mbf{a}_2^{eff} = -\mbf{a}_2 + \mbf{a}_3 & \nonumber \\  
~~~~~~~~~\mbf{a}_{S} =  \mbf{a}_3  & \label{2Dstackingvectors}  
\eeqa 
%\end{widetext}
%-----------------------------------------------------------------------------------------------------------
In all the figures in the text depicting data for the slab geometry we use a set of rotated axes to denote the
high symmetry points. For all the slab geometries the new x-axis is along $\mbf{a}_1^{eff}$ and the new z-axis,
is along the termination direction. For any slab geometry we truncate the lattice along the termination direction
without any further deletion of sites at the end. Thus for $N_l$ layers we have $4\times N_l$ bands in each case.
The lattice in the three slab orientations and the surface Brillouin zones associated with the three orientations
have been presented in Fig. \ref{BZ_GS_slabs_SBZ_fig}.
%-----------------------------------------------------------------------------------------------------------

%-----------------------------------------------------------------------------------------------------------
\ack
VRC acknowledges funding from the Department of Atomic Energy, India under the project number 12-R\&D-NIS-5.00-0100.
VVJ and VRC thank Mavani Himanshu Arvindbhai and Haraprasad Dhal for collaboration on closely related projects
pertaining to the effects of a magnetic field on the spin wave spectra of the model discussed here.

%-----------------------------------------------------------------------------------------------------------

\section*{References}
\bibliography{references_JPCM_post_review.bib}
%-----------------------------------------------------------------------------------------------------------
%-----------------------------------------------------------------------------------------------------------
%-------------------------------------------------------------------------------------------------------

\end{document}